\documentclass[twocolumn,floatfix,superscriptaddress,a4paper,showpacs,showkeys,aps]{revtex4-2}
\usepackage{epsfig}
 \usepackage{booktabs}
\usepackage{longtable}
\usepackage{latexsym}
\usepackage[colorlinks=true,linktocpage=true,linkcolor=blue,citecolor=blue,allcolors=blue]{hyperref}
\usepackage{url}
\usepackage[utf8]{inputenc}
\usepackage{enumerate}
\usepackage{color}
\usepackage{xcolor}
\usepackage{setspace}
\usepackage{amsmath}
\usepackage{amssymb}
\usepackage[english]{babel}
\usepackage{url}
\usepackage{appendix}
\begin{document}

 \title{QCD Equation of State of Dense Nuclear Matter \\
from a Bayesian Analysis of Heavy-Ion Collision Data}

\author{Manjunath Omana Kuttan}
\email{manjunath@fias.uni-frankfurt.de}
\affiliation{Frankfurt Institute for Advanced Studies, Ruth-Moufang-Str. 1, D-60438 Frankfurt am Main, Germany}
\affiliation{Institut f\"{u}r Theoretische Physik, Goethe Universit\"{a}t Frankfurt, Max-von-Laue-Str. 1, D-60438 Frankfurt am Main, Germany \looseness=-1}
\affiliation{Xidian-FIAS International Joint Research Center, Giersch Science Center, D-60438 Frankfurt am Main, Germany \looseness=-1}

\author{Jan Steinheimer}
\email{steinheimer@fias.uni-frankfurt.de}
\affiliation{Frankfurt Institute for Advanced Studies, Ruth-Moufang-Str. 1, D-60438 Frankfurt am Main, Germany}

\author{Kai Zhou}
\email{zhou@fias.uni-frankfurt.de}
\affiliation{Frankfurt Institute for Advanced Studies, Ruth-Moufang-Str. 1, D-60438 Frankfurt am Main, Germany}

\author{Horst Stoecker}
\email{stoecker@fias.uni-frankfurt.de}
\affiliation{Frankfurt Institute for Advanced Studies, Ruth-Moufang-Str. 1, D-60438 Frankfurt am Main, Germany}
\affiliation{Institut f\"{u}r Theoretische Physik, Goethe Universit\"{a}t Frankfurt, Max-von-Laue-Str. 1, D-60438 Frankfurt am Main, Germany \looseness=-1}
\affiliation{GSI Helmholtzzentrum f\"ur Schwerionenforschung GmbH, Planckstr. 1, D-64291 Darmstadt, Germany}

\date{\today}
\begin{abstract}
Bayesian methods are used to constrain the density dependence of the QCD Equation of State (EoS) for dense nuclear matter using the data of mean transverse kinetic energy and elliptic flow of protons from heavy ion collisions (HIC), in the beam energy range $\sqrt{s_{\mathrm{NN}}}=2-10~GeV$. The analysis yields tight constraints on the density dependent EoS up to 4 times the nuclear saturation density. The extracted EoS yields good agreement with other observables measured in HIC experiments and constraints from astrophysical observations both of which were not used in the inference. The sensitivity of inference to the choice of observables is also discussed.
\end{abstract}

\maketitle
The properties of dense and hot nuclear matter, governed by the strong interaction under quantum chromo dynamics (QCD), is an unresolved, widely studied topic in high energy nuclear physics. First principle lattice QCD studies, at vanishing and small baryon chemical potential, predict a smooth crossover transition from a hot gas of hadronic resonances to a chirally restored phase of strongly interacting quarks and gluons \cite{Aoki:2006we, Borsanyi:2010cj}. However, at high net baryon density i.e., large chemical potential, direct lattice QCD simulations are at present not available due to the fermionic sign problem \cite{Guenther:2017hnx}. Therefore, QCD motivated effective models as well as direct experimental evidence are employed to search for structures in the QCD phase diagram such as a conjectured first or second order phase transition and a corresponding critical endpoint \cite{Stephanov:2008qz,Stephanov:2011pb,Bluhm:2020mpc}. Diverse signals had been suggested over the last decades \cite{Stoecker:1986ci,Hofmann:1976dy,Stoecker:2004qu,Stephanov:1998dy,Hatta:2003wn}, but a conclusive picture has not emerged yet due to lack of systematic studies to relate all possible signals to an underlying dynamical description of the system, both consistently and quantitatively.

Recently, both machine learning and Bayesian inference methods have been employed to resolve this lack of unbiased quantitative studies. A Bayesian analysis has shown that the hadronic flow data in ultra relativistic heavy-ion collisions at the LHC and RHIC favors an EoS similar to that calculated from lattice QCD at vanishing baryon density \cite{Pratt:2015zsa}. In the high density range where lattice QCD calculations are not available, deep learning models are able to distinguish scenarios with and without a phase transition using the final state hadron spectra \cite{Pang:2016vdc, Steinheimer:2019iso, Du:2019civ, Jiang:2021gsw, OmanaKuttan:2020btb}.

This work presents a Bayesian method to constrain quantitatively the high net baryon density EoS from data of intermediate beam energy heavy-ion collisions. A recent study has attempted such an analysis by a rough, piecewise constant speed of sound parameterization of the high density EoS \cite{Oliinychenko:2022uvy}. In this study, a more flexible parameterization of the density dependence of the EoS is used in a model which can incorporate this density dependent EoS in a consistent way and then make direct predictions for different observables.

In this work, the dynamic evolution of heavy-ion collisions is entirely described by the microscopic Ultrarelativistic Quantum Molecular Dynamics (UrQMD) model \cite{Bass:1998ca,Bleicher:1999xi} which is augmented by a density dependent EoS. This approach describes the whole system evolution consistently within one model. No parameters besides the EoS itself are varied here. 

UrQMD is based on the propagation, binary scattering and decay of hadrons and their resonances. The density dependent EoS used in this model is realized  through an effective density dependent potential entering in the non-relativistic Quantum Molecular Dynamics (QMD) \cite{Aichelin:1986wa,Hartnack:1989sd,Stoecker:1986ci} equations of motions,
\begin{eqnarray}\label{motion}\dot{\textbf{r}}_{i}=\frac{\partial \mathrm{\bf{H}}}{\partial\textbf{p}_{i}},\quad \dot{\textbf{p}}_{i}=-\frac{\partial \mathrm{\bf{H}} }{\partial \textbf{r}_{i}}.\end{eqnarray}

Here $ \mathrm{\bf{H}} = \sum_i H_i$ is the total Hamiltonian of the system  including the kinetic energy and the total potential energy
${\mathrm{\bf{V}}=\sum_i V_i \equiv \sum_i V\big(n_B(r_i)\big)}$.
The equations of motion are solved given the potential energy $V$, which is related to the pressure in a straightforward manner \cite{Steinheimer:2022gqb}.

\begin{eqnarray} 
P(n_B) & = & P_{\rm id}(n_B) + \int_0^{n_B} n' \frac{\partial U(n')}{\partial n'}dn'\,\label{potential-to-eos}
\end{eqnarray}

Here, $ P_{\rm id}(n_B)$ the pressure of an ideal Fermi gas of baryons and $U(n_B)=\frac{\partial \big(n_B  \cdot V(n_B)\big)}{\partial n_B}$ is the single particle potential. Evidently, the potential energy is directly related to the EoS and therefore the terms potential energy and EoS are interchangeably used in this letter.

This model assumes that only baryons are directly affected by the potential interaction~\footnote{This simplification can be supported by the fact that at the beam energies under investigation, the EoS is dominated by the contribution from baryons}. A much more detailed description of the implementation of the density dependent potential can be found in \cite{OmanaKuttan:2022the, Steinheimer:2022gqb}. Note that this method does yield for bulk matter properties, strikingly similar results as the relativistic hydrodynamics simulations when the same EoS is used \cite{OmanaKuttan:2022the}.

To constrain the EoS from data, a robust and flexible parameterization for the density dependence of the potential energy that is capable of constructing physical equations of state (EOSs)  is necessary. For densities below twice the nuclear saturation density ($n_0$), the EoS is reasonably constrained by the QCD chiral effective field theory (EFT) calculations \cite{Tews:2012fj, Drischler:2017wtt}, data on nuclear incompressibility \cite{Wang:2018hsw}, flow measurements at moderate beam energies \cite{Danielewicz:2002pu,Kruse:1985hy,Molitoris:1985df,Stoecker:1986ci} and Bayesian analysis of both neutron star obervations and low energy heavy-ion collisions \cite{Huth:2021bsp}. This work focuses on the high density EoS, particularly on the range 2$n_0$- 6$n_0$, which is not well understood yet. Therefore, the potential energy $V(n_B)$ is fixed for densities up to $2n_0$ by using the Chiral Mean Field (CMF) model-fit to nuclear matter properties and flow data in the low beam energy region \cite{Steinheimer:2022gqb}. For densities above $2n_0$, the potential energy per baryon $V$ is parameterized by a seventh degree polynomial:
\begin{equation}\label{eq:poly}V(n_B)=\sum_{i=1}^{7}\theta_i \left(\frac{n_B}{n_0}-2\right)^i +h \end{equation}
where $h$=-22.07~MeV is set to ensure that the potential energy is a continuous function at $2n_0$.

This work constrains the parameters $\theta_i$ and thus the EoS, via Bayesian inference using the elliptic flow $v_2$ and the mean transverse kinetic energy $\left\langle m_T \right \rangle - m_0 $ of mid rapidity protons in Au-Au collisions at beam energy $\sqrt{s_{\mathrm{NN}}} \approx 2 - 10$ GeV. The ${v_2}$ data are from mid-central collisions at $\sqrt{s_{\mathrm{NN}}}=$ 2.24, 2.32, 2.4, 2.42, 2.51, 3.0, 3.32, 3.84, 4.23 and 4.72 GeV \cite{E895:1999ldn,CERES:2002eru,FOPI:2004bfz,STAR:2012och,STAR:2020dav,HADES:2020lob,STAR:2021yiu}  and the $\left\langle m_T \right \rangle - m_0 $ data are from central collisions at $\sqrt{s_{\mathrm{NN}}}=$ 3.83, 4.29, 6.27, 7.7 and 8.86 GeV \cite{E802:1999hit,NA49:2006gaj,STAR:2017sal}. Important, sensitive observables such as the directed flow \cite{Stoecker:2004qu, Nara:2016hbg} are then used to cross check the so extracted EoS. The choice of proton observables (as proxy to baryons) is due to the fact that interesting features in the EoS at high baryon density and moderate temperatures are dominated by the interactions between baryons and protons form the most abundant hadron species, actually measured in experiments, for beam energies considered in present work. Further details on the choice of data and calculation of flow observables are given in  appendix \ref{a:data}, which includes Ref. \cite{Reichert:2022gqe}. 

The experimental data $\textbf{D}=\{v_2^{exp}, \left\langle m_T \right \rangle^{exp} - m_0 \}$ are used to constrain the parameters of the model $\boldsymbol\theta = \{\theta_1, \theta_2,...,\theta_7 \}$ by using the Bayes theorem, given by 
\begin{equation}\label{eq:bayes}P(\boldsymbol\theta|\textbf{D})\propto P(\textbf{D}|\boldsymbol\theta) P(\boldsymbol\theta).\end{equation}
Here $P(\boldsymbol\theta)$ is the prior distribution, encoding our prior knowledge on the parameters while $P(\textbf{D}|\boldsymbol\theta)$ is the likelihood for a given set of parameters which dictates how well the parameters describe the observed data. Finally, $P(\boldsymbol\theta|\textbf{D})$ is the desired posterior which codifies the updated knowledge on the parameters $\boldsymbol\theta$ after encountering the experimental evidence $\textbf{D}$.

The objective is to construct the joint posterior distribution for the 7 polynomial coefficients ($\boldsymbol\theta$) based on experimental observations, for which Markov Chain Monte Carlo (MCMC) sampling methods are used. For an arbitrary parameter set, the relative posterior probability up to an unknown normalisation factor is simply given by the prior probability as weighted by its likelihood. To evaluate the likelihood for a parameter set, the $v_2$ and the $\left\langle m_T \right \rangle - m_0 $ observables need to be calculated by UrQMD. The MCMC method then constructs the posterior distribution by exploring the high dimensional parameter space based on numerous such likelihood evaluations. This requires numerous computationally intensive UrQMD simulations which would need unfeasible computational resources. Hence, Gaussian Process (GP) models are trained as fast surrogate emulators for the UrQMD model, to interpolate simulation results in the parameter space \cite{doi:10.1080/01621459.1991.10475138,Bernhard:2019bmu,Novak:2013bqa,Pratt:2015zsa}. Cuts in
rapidity and centrality that align with that of the experiments are applied on UrQMD data to create training data for the GP models. The constraints applied to generate the physical EoSs to train the models, the performance of the GP models and other technical details can be found in  appendix \ref{sec2}.

The prior on the parameter sets is chosen as Gaussian distributions with means and variances evaluated under physical constraints. More details on the choice of the priors are given in  appendix \ref{secprior}. The log-likelihood is evaluated using
uncertainties from both the experiment and from the GP model. The prior, together with the trained GP-emulator, experimental observations and the likelihood function are used for the MCMC sampling by employing the DeMetropolisZ \cite{demetro,ter2008differential} algorithm from PyMC v4.0 \cite{pymc}.

\begin{figure}[t]
   \includegraphics[width=0.5\textwidth,keepaspectratio]{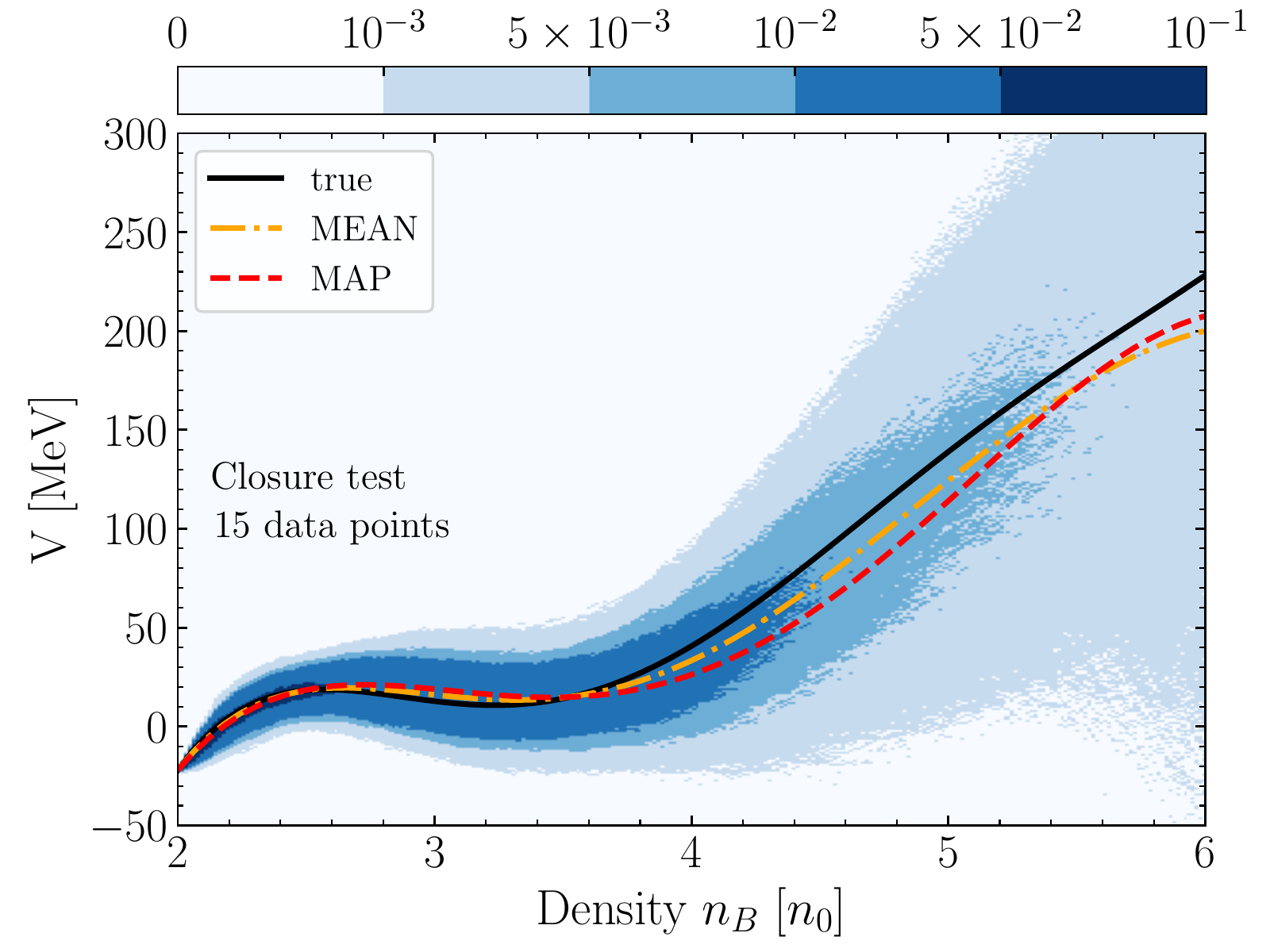}
   \caption{(Color online) Visualisation of the sampled posterior in the closure test. The color represents the probability for the potential at a given density. The `ground-truth' EoS used for generating the observations is plotted as black solid line. The red dashed and orange dot-dashed curves are the MAP and MEAN EoS for the posterior.}
   \label{closure}
 \end{figure}

\emph{Closure tests.} In order to verify the performance of the Bayesian inference method described above, two closure tests are performed. The first test involves constructing the posterior using $v_2$ and $\left\langle m_{T} \right\rangle- m_0$, simulated with the experimental uncertainties from UrQMD for a specific but randomly chosen EoS. The inference results are then compared to the known `ground-truth'. Figure \ref{closure} shows the posterior constructed in one such test for a random input potential. The black curve in the plot is the `ground-truth' input potential while the color contours represent the reconstructed probability density for a  given value of the potential $V(n_b)$. Two specific estimates of the `ground-truth' potential are highlighted in the figure besides the posterior distribution of the potential.
These are the Maximum A Posteriori (MAP) estimate, which represents the mode of the posterior distribution as evaluated via MCMC and the `MEAN' estimate as calculated by averaging the values of the sampled potentials at different densities. The comparison of the MAP and the MEAN curves with the `ground-truth' shows that the reconstruction results from the Bayesian Inference are centered around the 'ground-truth' EoS and the sampling converges indeed to the true posterior. From the spread of the posterior it can be seen that the EoS in the closure test is well constrained up to densities ~$4n_0$ for the observables used in the present study. For densities from ~$4n_0$ up to $6n_0$ the generated EoSs have larger uncertainties. However, the mean potentials follow closely the true potential. 

The second closure test is done in order to determine the sensitivity of the inference to the choice of the observational data. Hence, the procedure is similar to the previous test, except that the $\left\langle m_{T} \right\rangle - m_0 $ values for $\sqrt{s_{\mathrm{NN}}}$= 3.83 and 4.29 GeV are not used in this test to estimate the posterior. When these two data points are excluded, the agreement of the `ground-truth' EoS with the MAP and MEAN estimates decreases considerably for densities greater than 4$n_0$. This indicates that these data points are crucial indeed for constraining the EoS at higher densities. Further details about these closure tests, and the sensitivity on excluding different data points, can be found in appendices \ref{secclose}, \ref{secpripos} and \ref{secsense}. There, also a comparison of the prior and posterior probability distributions is shown to highlight the actual information gain obtained through the Bayesian inference.

\begin{figure}[t]
   \includegraphics[width=0.5\textwidth,keepaspectratio]{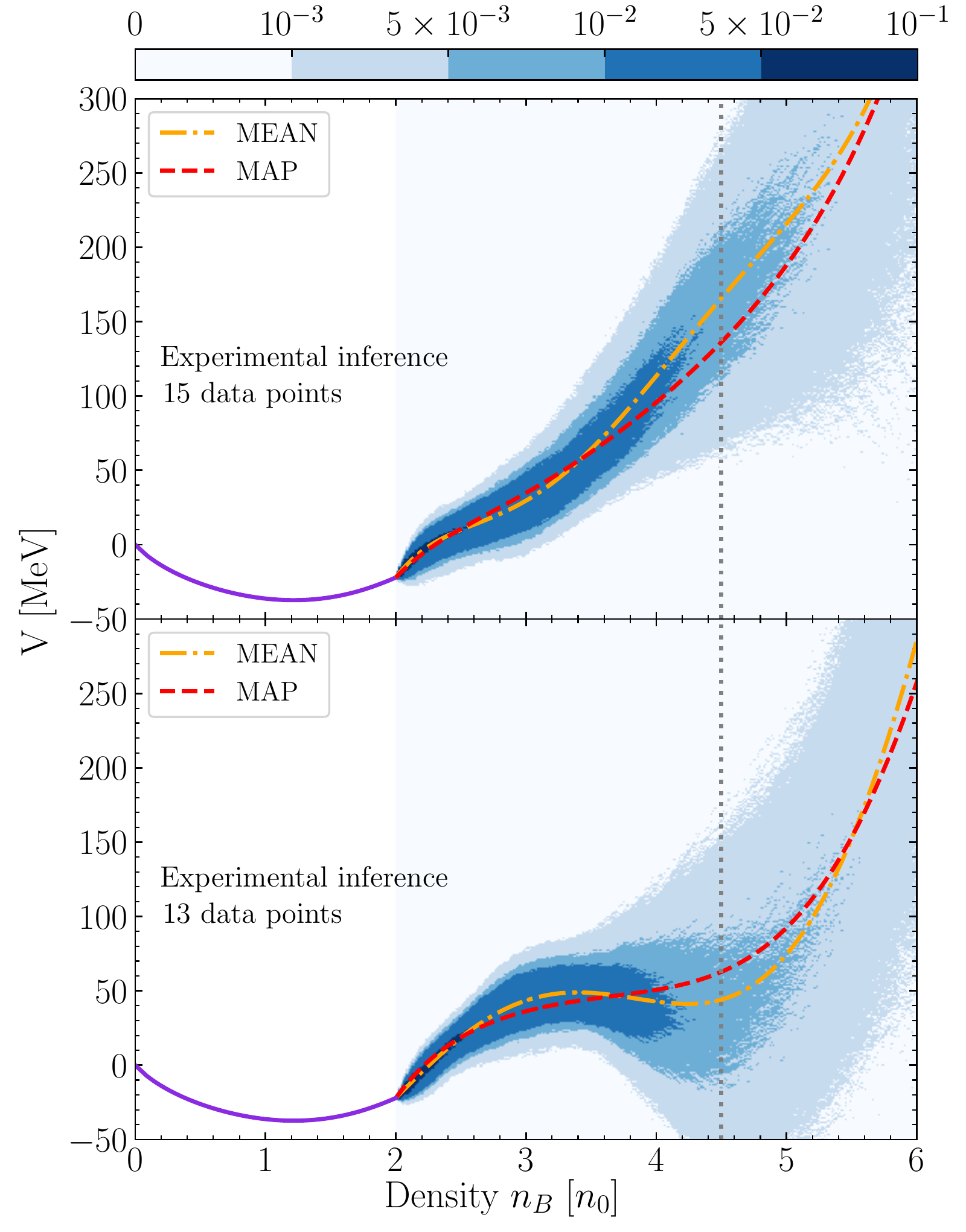}
   \caption{(Color online) Posterior distribution for the EoS inferred using experimental observations of $v_2$ and $\left\langle m_{T} \right\rangle -m_0$. The top figure is the posterior when all 15 data points were used while the bottom figure is obtained without using the $\left\langle m_{T} \right\rangle - m_0 $ values for $\sqrt{s_{\mathrm{NN}}}$= 3.83 and 4.29 GeV. The MAP and MEAN EoSs in both cases are plotted in red dashed and orange dot-dashed curves respectively. The vertical, grey line depicts the highest average central compression reached in collisions at $\sqrt{s_{\mathrm{NN}}}$=9 GeV. The CMF EoS is plotted in violet for density below 2$n_0$.}
   \label{realdata}
 \end{figure}

\emph{Results based on experimental data:}
The results of sampling the posteriors by using experimental data, for the two cases, with and without the $\left\langle m_{T} \right\rangle - m_0 $ values at $\sqrt{s_{\mathrm{NN}}}$= 3.83 and 4.29 GeV, are shown in figure \ref{realdata}. The upper panel corresponds to using 15 experimental data points while the lower panel shows the results without the two $\left\langle m_{T} \right\rangle - m_0 $ values. The data as used in this paper do well constrain the EoS, for densities from $2n_0$ to $4n_0$. However, beyond $4n_0$, the sampled potentials have a large uncertainty and the variance is significantly larger for the posterior extracted from 13 data points. Beyond densities of about $3n_0$, the posterior extracted using 13 data points differs significantly from the posterior extracted using all 15 points. This is quite different from our closure tests, where the extracted MAP and MEAN curves did not depend strongly on the choice of the data points used. This indicates a possible tension within the data in the context of the model used.

\begin{figure}[t]
   \includegraphics[width=0.5\textwidth,keepaspectratio]{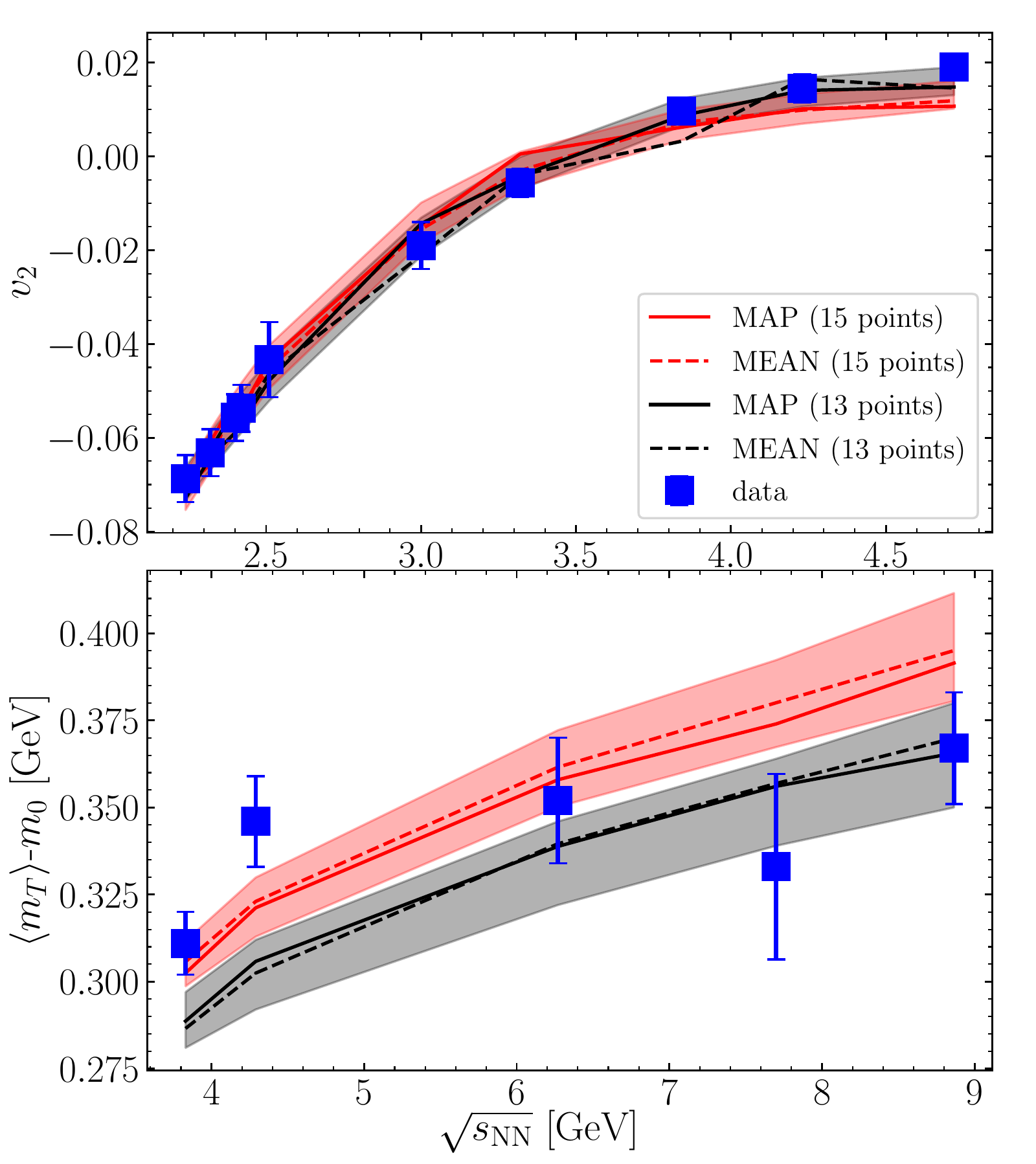}
   \caption{(Color online) $v_2$ and $\left\langle m_{T} \right\rangle-m_0$ values from UrQMD using the MEAN and MAP EoS as extracted from measured data. The observables for both MAP and MEAN EoSs, extracted by using all 15 data points are shown as solid and dashed red lines respectively, while those generated using only the 13 data points are shown as solid and dashed black lines respectively. The experimental data are shown as blue squares. The uncertainty bands correspond to a 68 \% credibility constraint constructed
from the posterior samples}
   \label{close2}
 \end{figure}

To understand this significant deviation which appears when only two data points are removed, the MAP and MEAN EoS resulting from the two scenarios are implemented into the UrQMD model to calculate the $v_2$ and the $\left\langle m_{T} \right\rangle - m_0 $ values which are then compared with the experimental data which were used to constrain them. Figure \ref{close2} shows the MAP and MEAN curves together with 1-sigma confidence intervals from the posterior. Both results, with different inputs, fit the $v_2$ data very well except for the small deviation at the high energies. The fit is slightly better when the $\left\langle m_{T} \right\rangle-m_0$ values at the lowest energies are removed. At the same time, using all data points results in larger $\left\langle m_{T}\right\rangle -m_0$ values for both the MAP and MEAN curves. The bands for $\left\langle m_{T} \right\rangle-m_0$ are much broader than the bands for $v_2$. Yet, the uncertainty bands clearly support the differences in the fit portrayed by the MEAN and MAP curves. The model encounters a tension between the $\left\langle m_{T} \right\rangle - m_0 $ and the $v_2$ data. This tension may either be due to a true tension within the experimental data, or due to a shortcoming of the theoretical model used to simulate both the  $\left\langle m_{T} \right\rangle - m_0 $ and the $v_2$ data at high beam energies for a given equation of state. It should also be noted that at higher beam energies the contributions from the mesonic degrees of freedom to the equation of state becomes more dominant which may make an explicitly temperature dependent equation of state necessary.

\begin{figure}[t]
   \includegraphics[width=0.5\textwidth,keepaspectratio]{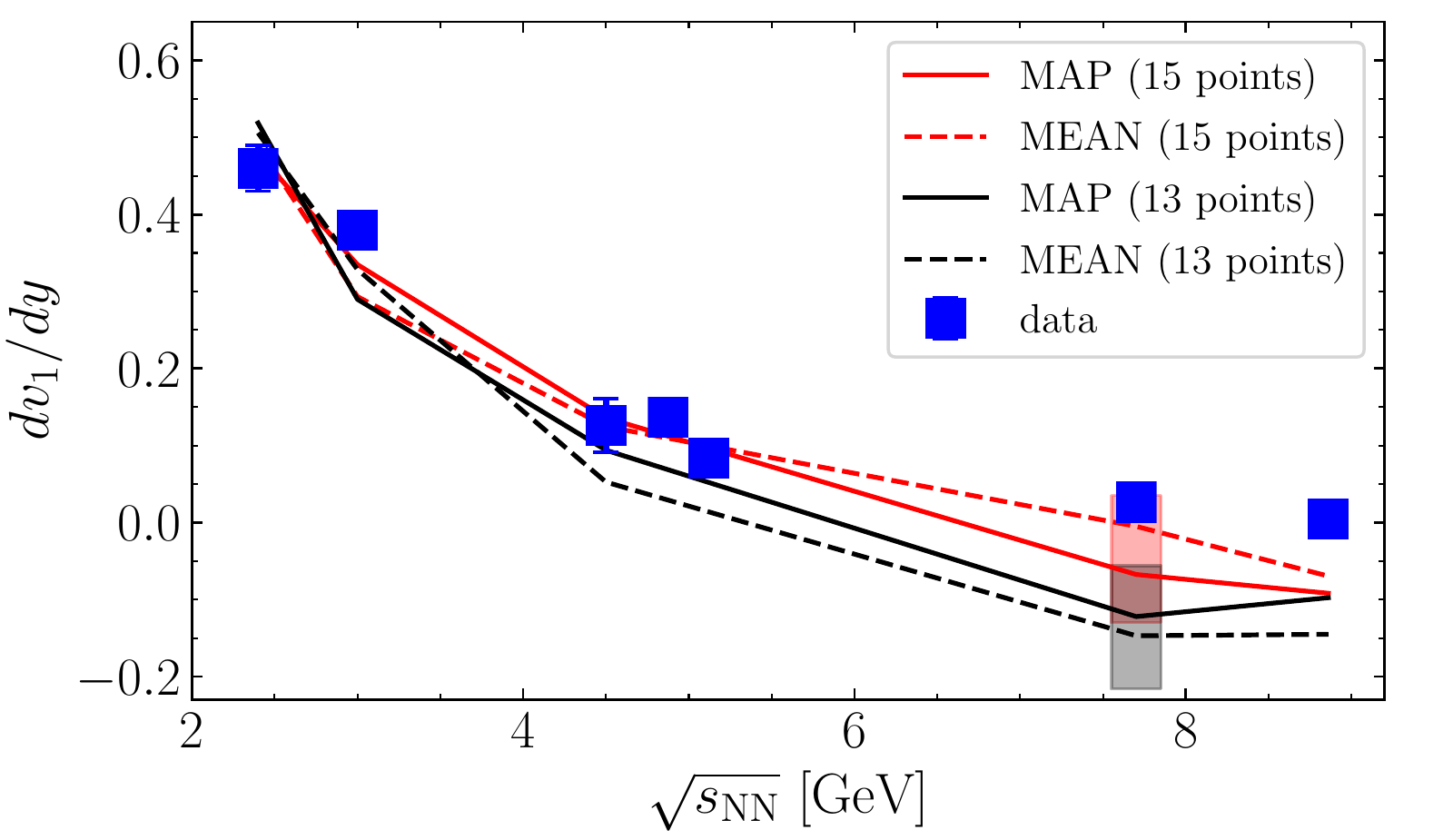}
   \caption{(Color online) Slope of the directed flow, $dv_1/dy$, of protons at mid rapidity. The experimental data \cite{STAR:2020dav,STAR:2021ozh,STAR:2021yiu,HADES:2020lob,HADES:2022osk,Kashirin:2020evw,E877:1996czs,NA49:2003njx}are shown as blue squares. The colored bars correspond to a 68\% credibility constraint constructed from the posterior samples. }
   \label{v1}
 \end{figure}

Finally, the extracted EoS can be tested using various observables like differential flow measurements (see appendix \ref{secdiff}, which include Refs. \cite{Savchuk:2022aev,Li:2022iil,Savchuk:2022msa,Reichert:2023eev,HADES:2022osk}) or different flow coefficients. The slope of the directed flow $dv_1/dy$ at mid rapidity are calculated using the reconstructed MEAN and MAP EoSs. The results together with available experimental data are shown in figure \ref{v1}. The $dv_1/dy$ prediction closely match the experimental data, especially at the higher energies, for the MEAN EoS extracted from all 15 data points. The 1-sigma confidence intervals are indicated as colored bars. It is shown only for one beam energy due to the high computational cost. It can be seen that at high energies, in the 13-points case, the prediction clearly undershoots the data while in the 15-points case, the experimental data lies at the
border of the 1-sigma band. The reconstructed EoSs for all other energies are consistent with the $dv_1/dy$ data though it was not used to constrain the EoSs.

 \begin{figure}[t]
   \includegraphics[width=0.5\textwidth,keepaspectratio]{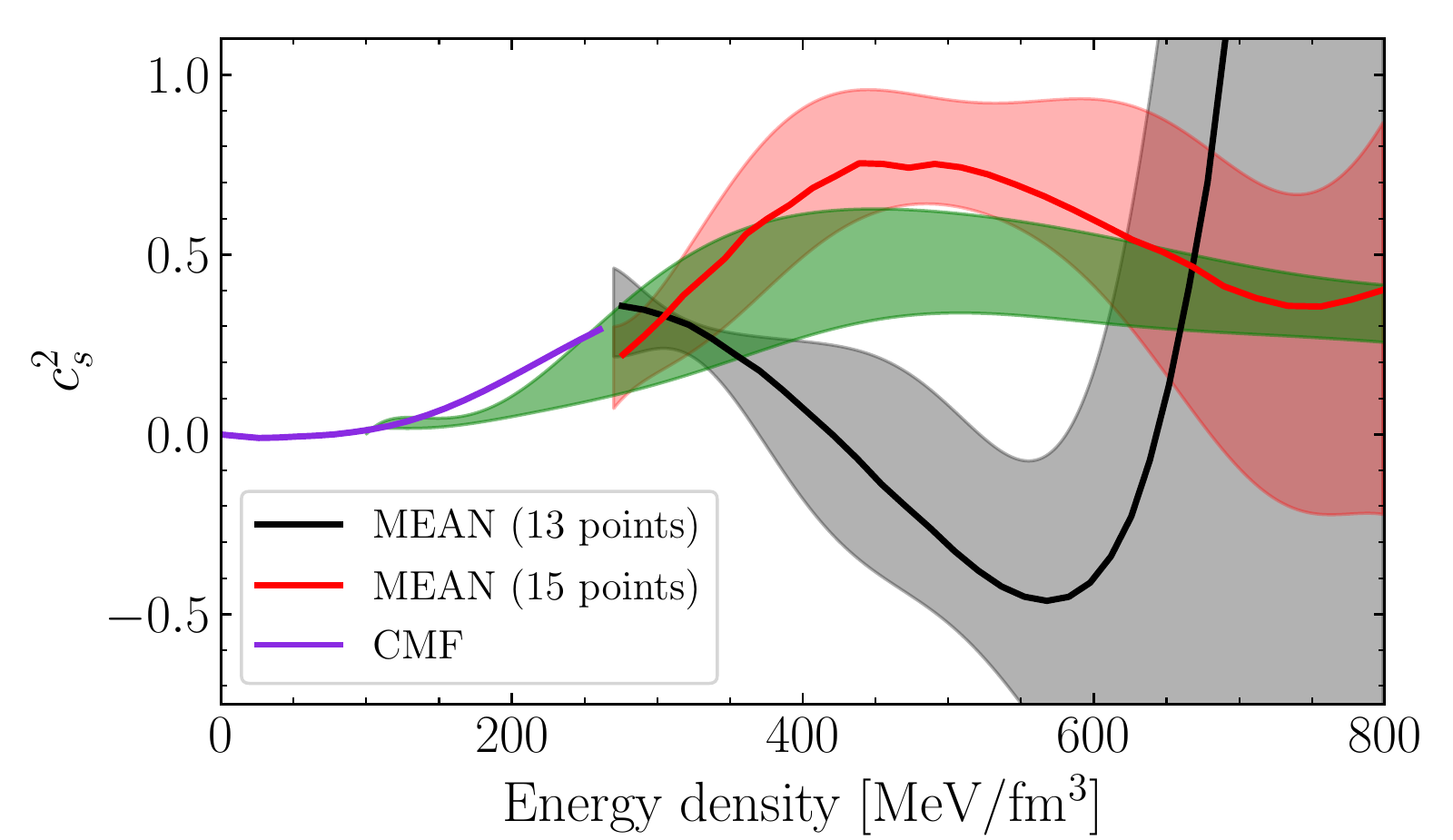}
   \caption{(Color online) Speed of sound squared $c_s^2$, at $T=0$, as a function of energy density. The $c_s^2$ for the MEAN EoS extracted from all data points are shown in red and those extracted from only 13 data points are shown in black. The constraints from astrophysical observations are shown as a green band. For energy densities up to 270 MeV/fm$^3$, the speed of sound from CMF is plotted as violet curve. The uncertainty bands correspond to a 68$\%$ credibility constraint from the inferred potential curves.
}
   \label{cs}
 \end{figure}
To relate the extracted high density EoS to constraints from astrophysical observations, the squared speed of sound ($c_s^2$) at $T=0$ is presented for the MEAN EoSs as a function of the energy density in Figure \ref{cs}, together with a contour which represents the constraints from recent Binary Neutron Star Merger (BNSM) observations~\cite{Altiparmak:2022bke} \footnote{Note, that even though the two systems have a different isospin fraction, the effect of the isospin composition is likely small at large densities}. The speed of sound, as the derivative of the pressure is very sensitive to even small variations of the potential energy. The $c_s^2$ values estimated from all data points show overall agreement with the $c_s^2$ constraints from astrophysical observations and predicts a rather stiff equation of state at least up to 4$n_0$. In particular, both the astrophysical constraints (see also \cite{Soma:2022vbb}) and the EoS inference in the present work gives a broad peak structure for $c_s^2$. This is compatible with recent functional renormalization group (FRG)~\cite{Leonhardt:2019fua} and conformality~\cite{Fujimoto:2022ohj} analyses. However, if only the 13 data points are used, the extracted speed of sound shows a drastic drop, consistent with a strong first order phase transition at high densities\cite{Hofmann:1976dy,Stoecker:2004qu}. This is consistent with the softening phenomenon observed for $\left\langle m_{T} \right\rangle -m_0 $ data shown in Figure.~\ref{close2}.  In order to give an estimate of the uncertainty on the speed of sound, we have calculated the speeds of sound for 100000 potentials which lie within the $68\%$ credibility interval of the coefficients, however excluding those which lead to acausal equations of state for densities below 4.5 n0.

\emph{Conclusion.} Bayesian inference can constrain the high density QCD EoS using experimental data on $v_2$ and $\left\langle m_{T} \right\rangle-m_0$ of protons. Such an analysis, based on HIC data, can verify the dense QCD matter properties extracted from neutron star observations and  complements astrophysical studies to extract the finite temperature EoS from BNSM merger signals as well as constrain its dependence on the symmetry energy. 

A parametrized density dependent potential is introduced in the UrQMD model used to train Gaussian Process models as fast emulators to perform the MCMC sampling. In this framework, the input potential can be well reconstructed from experimental HIC observables available already now from experimental measurements. The experimental data constrain the posterior constructed in our method for the EoS, for densities up to $4n_0$. However, beyond $3n_0$, the shape of the posterior depends on the choice of observables used. As a result, the speed of sound extracted for these posteriors exhibit obvious differences. The EoS extracted using all available data points is in good agreement with the constraints from BNSMs with a stiff EoS for densities up to 4$n_0$ and without a phase transition. A cross check is performed with the extracted potentials by calculating the slope of the directed flow.  Here, a MEAN potential extracted from all 15 data-points gives the best, consistent description of all available data. The inferences encounter a tension in the measurements of $\left\langle m_{T} \right\rangle -m_0 $ and $v_2$ at a collision energy of $\approx$4 GeV. This could indicate large uncertainties in the measurements, or alternatively the inability of the underlying model to describe the observables with a given input EoS. Note, that the data are from different experiments that have been conducted during different time periods. The differences in the acceptances, resolutions, statistics and even analysis methods of experimental data makes it difficult for us to pin down the exact sources of these effects.

Tighter constraints and fully conclusive statements on the EoS beyond density $3n_0$ require accurate, high statistics data in the whole beam energy range of 2-10 GeV which will hopefully be provided by the beam energy scan program of STAR-FXT at RHIC, the upcoming CBM experiment at FAIR and future experiments at HIAF and NICA. It is noted that, when approaching higher beam energies, which would be important in extending the constraints to higher temperatures and/or densities, the currently used transport model needs to incorporate further finite-temperature and possible partonic matter effects together with relativistic corrections, which we leave for future studies. Further effort should be put into the development and improvement of the theoretical models to consistently incorporate different density dependent EoSs for the study of systematic uncertainties \cite{Sorensen:2023zkk}. In future, the presented method can also be extended to include more parameters of the model as free parameters for the Bayesian inference, which would also require more and precise input data.
In addition, other observables such as the higher order flow coefficients and $v_1$ can be incorporated into the Bayesian analysis, if permitted by computational constraints, for a more comprehensional constraint of the EoS in the future. 

\textbf{Acknowledgment. } The authors thank Volker Koch, Behruz Kardan and Shuzhe Shi for insightful discussions. This work is supported by the Helmholtzzentrum f\"ur Schwerionenforschung GSI (M.OK), the BMBF under the ErUM-Data and KISS projects (M. O. K, K. Z) project (M.O.K, K.Z), the SAMSON AG (M.O.K, J.S, K.Z) and the Walter Greiner Gesellschaft zur F\"orderung der physikalischen Grundlagenforschung e.V. through the Judah M. Eisenberg Laureatus Chair at the Goethe Universit\"at Frankfurt am Main (H.S).

\bibliography{Bl_EoS}
\bibliographystyle{apsrev4-1}

\appendix
\section{The data}
\label{a:data}
The study uses measurements of the elliptic flow $v_2$ of protons at ten different beam energies and transverse kinetic energy $\left\langle m_T \right \rangle - m_0 $ of protons at five different beam energies to constrain the EoS. The experimental measurements of $v_2$ and $\left\langle m_T \right \rangle - m_0 $ are from mid-central collisions and central collisions respectively and only mid-rapidity protons are considered. To calculate these observables from the UrQMD model, similar cuts in rapidity and centrality are applied. 

The $v_2$ is calculated from UrQMD data as,
\begin{equation}
\label{eq:v2}
v_{2}=\left\langle \frac{P_{x}^{2}-P_{y}^{2}}{P_{x}^{2}+P_{y}^{2}} \right \rangle 
\end{equation}
where the momenta are defined with respect to the reaction plane of the model.
At low beam energies event-plane or cumulant methods to extract the elliptic and directed flow are usually not used due to the significant interactions between the spectators and the participant region which leads to the negative v2 and strong directed flow. For a more detailed discussion on the flow correlations at SIS18 energies we refer to \cite{Reichert:2022gqe}. Thus, experiments in this energy range usually have dedicated detectors to determine the actual reaction plane of the collisions. In our analysis we also calculate the flow with respect to the reaction plane. In this way, non-flow effects, e.g., from multi-baryon correlations not related to the collective flow, are not included in the analysis.

Both $v_2$ and $\left\langle m_T \right \rangle - m_0 $ are calculated for protons at mid-rapidity ($\mid y/y_b \mid < 0.1$, where $y_b$ is the beam rapidity in the center of mass frame). For a given EoS, to calculate $v_2$ and  $\left\langle m_T \right \rangle - m_0 $  with errors similar to the experimental error, 12000 mid-central ($5< b < 8.3 $ fm) and 1000 central collision events ($ 0 < b < 3.4 $ fm ) respectively, are used.

The choice of proton observables is clearly motivated by the fact that in the considered beam energies, the dynamics is dominated by the baryons and pions 'feel' the effect of the density dependent potential only indirectly through e.g. the baryonic resonance decays. The elliptic flow was selected as there exists a vast amount of high precision measurements with relatively small systematic uncertainties. On the other hand, the transverse kinetic energy measurements are only available for few beam energies. However, the transverse kinetic energy can be calculated with precision similar to that in the experimental data using fewer events (about 1000 events). This makes the transverse kinetic energy measurements a good choice of observable in addition to $v_2$ for constraining the EoS. 

\begin{figure}[t]   
   \includegraphics[width=0.5\textwidth,keepaspectratio]{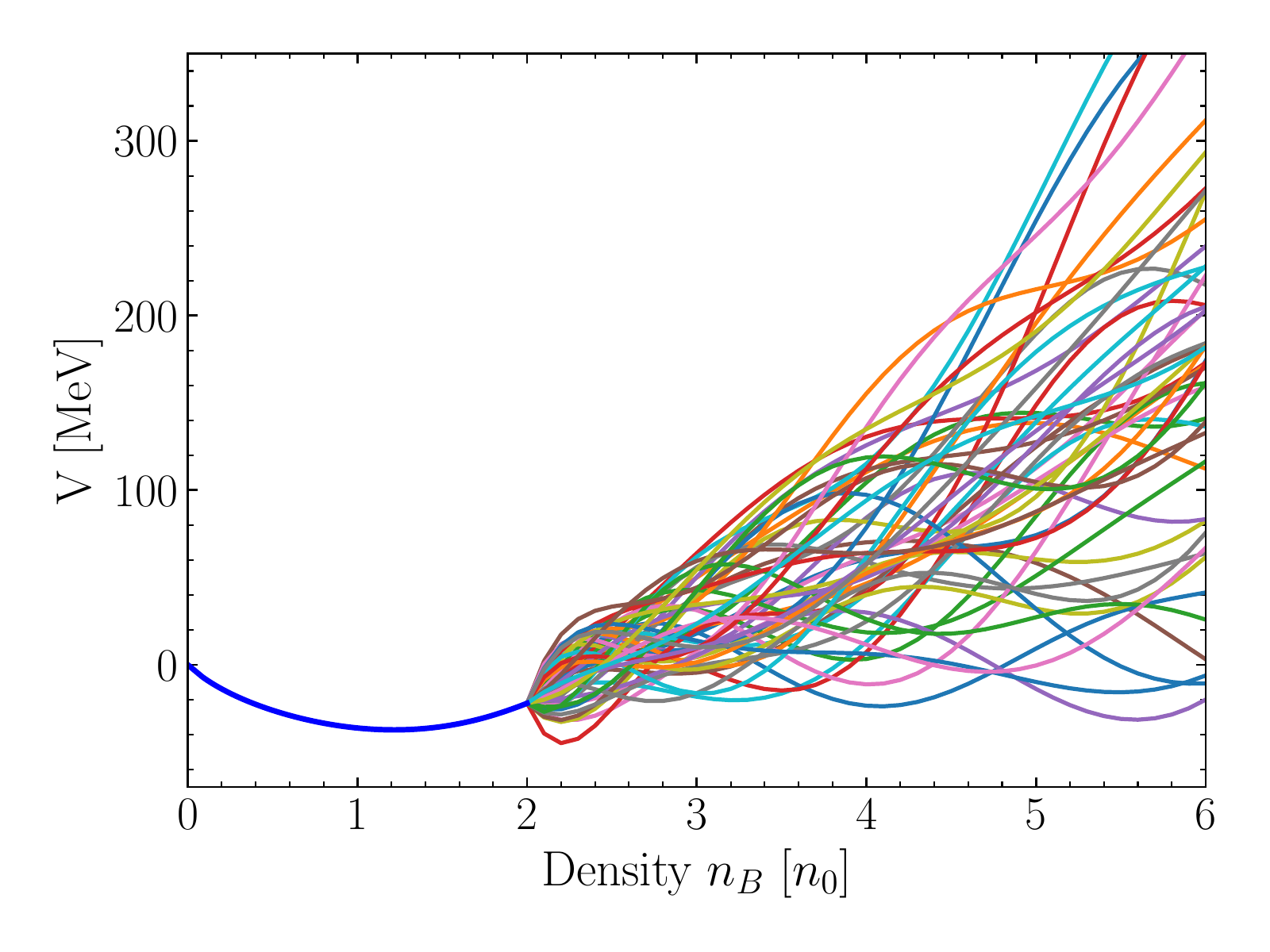}
   \caption{(Color online) Visualisation of some of the EoS used in training the Gaussian Process models. The dark blue line is the CMF EoS and the lines starting from $2n_0$ are the different polynomial EoS. The CMF and Polynomial EoS are forced to match at $2n_0$. The plot reveals the flexibility of the polynomial parameterisation in constructing different EoSs.}
   \label{eosparam}
 \end{figure}

\section{Training the Gaussian Process models}\label{sec2}
The Gaussian Process models used in this study take the 7 polynomial coefficients as input and predict the $v_2$ or $\left\langle m_T \right \rangle - m_0$ observables. Figure \ref{eosparam} shows a set of example curves that were randomly generated using the polynomial parameterization of the EoS. Such EoSs are used as input to UrQMD for calculating the $v_2$ and $\left\langle m_T \right \rangle - m_0$ observables. To avoid unrealistic EoSs in the training data, several constraints are applied for the potential functions for densities $2-8~n_0 $. A lower limit of about -40 MeV is set for the value of the potential to prevent the formation of a second bound state while the upper limit is set to be atmost 50 MeV higher than the value of a hard Skyrme EoS at any given density to avoid superluminal EoSs. Moreover, the potentials that are generated for training the GP models are constrained to have a derivative $dV/dn_B$ approximately within [-350, 450] MeV/$n_0$  for densities $2-8~n_0 $ to prevent the potential from fluctuating too strongly. Note, that these constrains are only used in generating the training data for the GP models and are not applied during the MCMC sampling.

\begin{figure}[t]
   \includegraphics[width=0.5\textwidth,keepaspectratio]{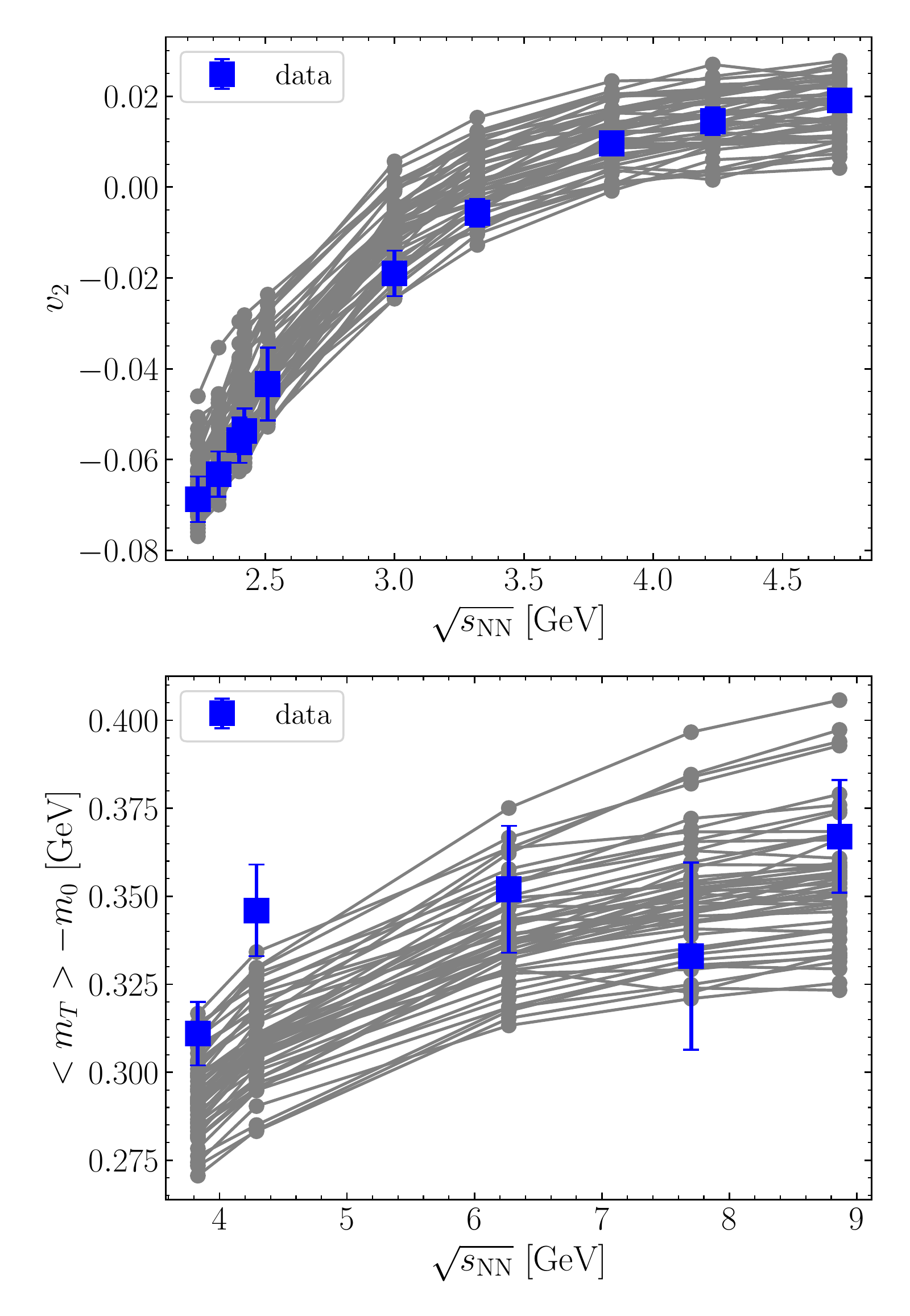}
   \caption{(Color online)Visualisation of the $v_2$ and  $\left\langle m_T \right \rangle - m_0$ for 50 random EoSs from the training data. The upper plot is the $v_2$ and the lower plot is the $\left\langle m_T \right \rangle - m_0$ as a function of $\sqrt{s_{\mathrm{NN}}}$. The experimental measurements are plotted in blue squares while the gray lines are from the training EoSs. }
   \label{data}
 \end{figure}

The simulated $v_2$ and $\left\langle m_T \right \rangle - m_0$ values for several random EoSs used for training the GP models are shown in figure \ref{data}. It is evident from the figure that our training data is diverse enough to cover a wide range of values for $v_2$ and $\left\langle m_T \right \rangle - m_0$ around the experimental observations. 

\begin{figure*}[]
   \includegraphics[width=0.91\textwidth,height=23cm,]{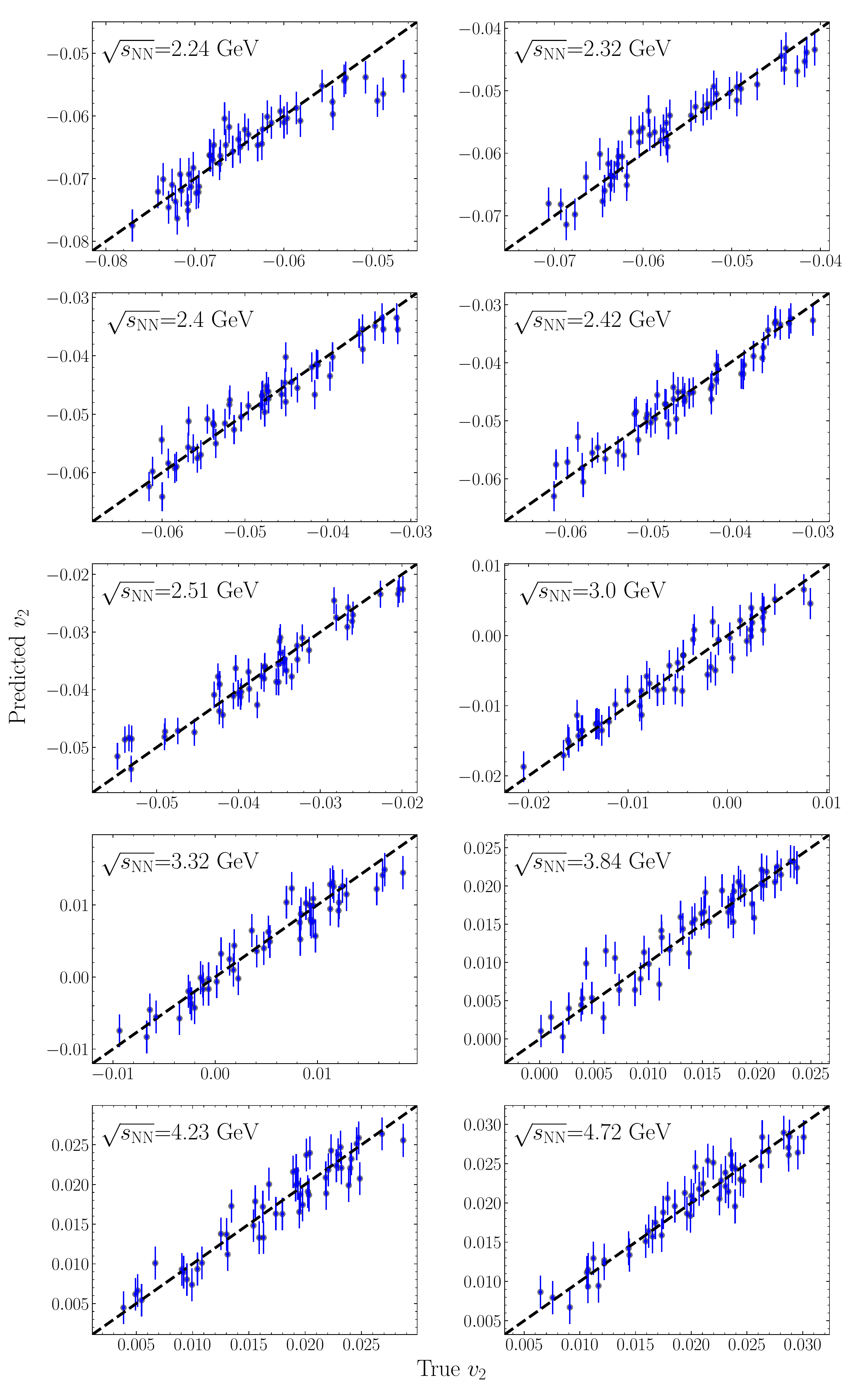}
   \caption{(Color online) Performance of the Gaussian Process models which predict the $v_2$ at different collision energies. The predictions for 50 different EoSs in the validation dataset are shown in blue while the error bar is the standard deviation of the prediction returned by the GP model. The true=predicted curve is shown as black, dashed line for reference.  }
   \label{gpv2}
 \end{figure*}

\begin{figure*}[]
   \includegraphics[width=0.9\textwidth,height=15.2 cm]{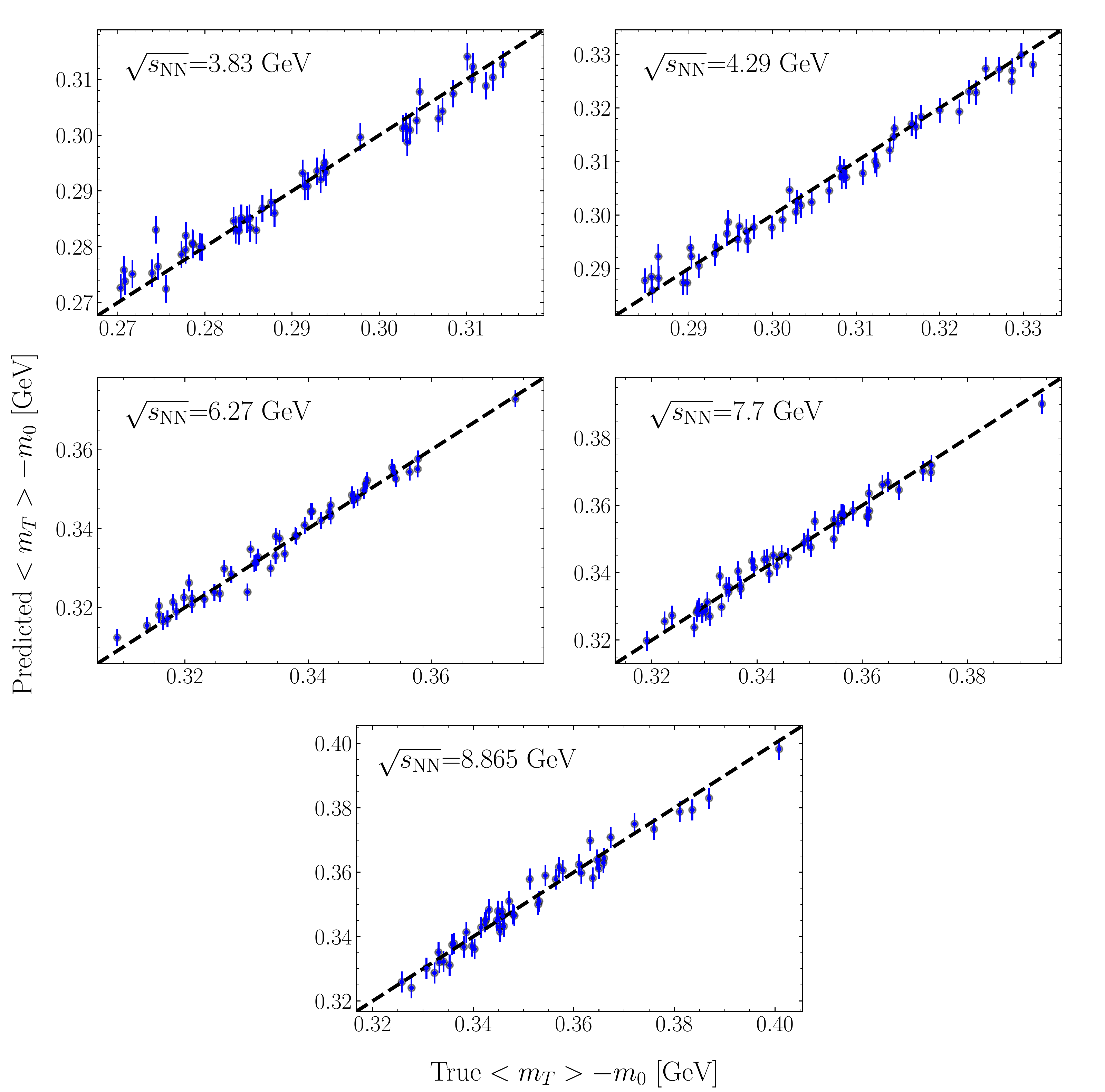}
   \caption{(Color online) Performance of the Gaussian Process models in predicting the $\left\langle m_T \right \rangle - m_0$ for 5 different collision energies. The predictions are shown in blue while the black, dashed line depicts the true= predicted curve.}
   \label{gpmmt}
 \end{figure*}

The GP emulators are trained on a set of 200 different parameter sets, each with a different high density EoS and the performance of these models is then validated on another 50 input parameter sets. 15 different GP models are trained, each one predicting one of the observables ($v_2$ for 10 collision energies + $\left\langle m_T \right \rangle - m_0$ for 5 collision energies). The trained GP models can be evaluated by comparing the GP predictions with the "true" results of UrQMD simulations. The performance of the GP models in predicting the $v_2$ and $\left\langle m_T \right \rangle - m_0$ observables for 50 different EoSs in the validation dataset are shown in figures \ref{gpv2} and \ref{gpmmt} respectively. As evident in these plots, the GP models can accurately predict the simulated observables, given the polynomial coefficients. Hence, the GP models can be used as fast emulators of UrQMD during the MCMC sampling. All the posterior distributions presented in this work are constructed by 4 different MCMC chains.  Each chain generates 25000 samples after 10000 tuning steps.

 \section{The prior}
\label{secprior}
In the following we will explain the choice of the prior distributions which is used as starting point of the Bayesian inference. Technically speaking, the prior distribution of parameters $\theta_i$ are chosen as Gaussian distributions whose means and variances are estimated from the randomly sampled EoSs, under physical constraints, used in the training of the Gaussian Process Emulators.  These constraints were introduced to ensure numerically stable results in training the GP models.
To create such a robust training dataset, different physics constraints were applied as discussed in appendix \ref{sec2}. These constraints eliminate some of the wildly fluctuating and superluminal EoSs from the training data. 

To ensure that the prior in the analysis is broad enough to reflect an a priori high degree of uncertainty (i.e., without introducing a bias) the mean and width of the distributions in the constraint GP training where used also in the prior. However, the polynomial coefficients $\theta_i$ resulting from these constraints, used to construct the prior distributions for the Bayesian inference, are then sampled independently and are thus not correlated as they would be in the GP model training. Thus, the priors for the Bayesian inference are much broader than the distributions used for the GP model training. The means and standard deviations of the Gaussian priors for the polynomial coefficients are shown in the table \ref{table_prior}.
\begin{table}[]
\centering
\begin{center}
\resizebox{0.8\linewidth}{!}{%
\begin{tabular}{c c c c c c c c}

\toprule
\toprule
 & $\theta_1$    & $\theta_2$    & $\theta_3$   & $\theta_4$      & $\theta_5$   &$\theta_6$   & $\theta_7$    \\ \midrule
$\mu$    & 77.5 & -78.7 & 65.4& -25.96 & 5.52 & -0.6& 0.03 \\
$\sigma$ & 150  & 450   & 450  & 225    & 55  & 7   & 0.3 \\\bottomrule
\bottomrule
\end{tabular}}
\end{center}
\caption{\label{table_prior} Means ($\mu$) and standard deviations ($\sigma$) of the Gaussian priors for the seven polynomial coefficients ($\theta_i$).}
\end{table}

\begin{figure*}[]
   \includegraphics[width=0.85\textwidth]{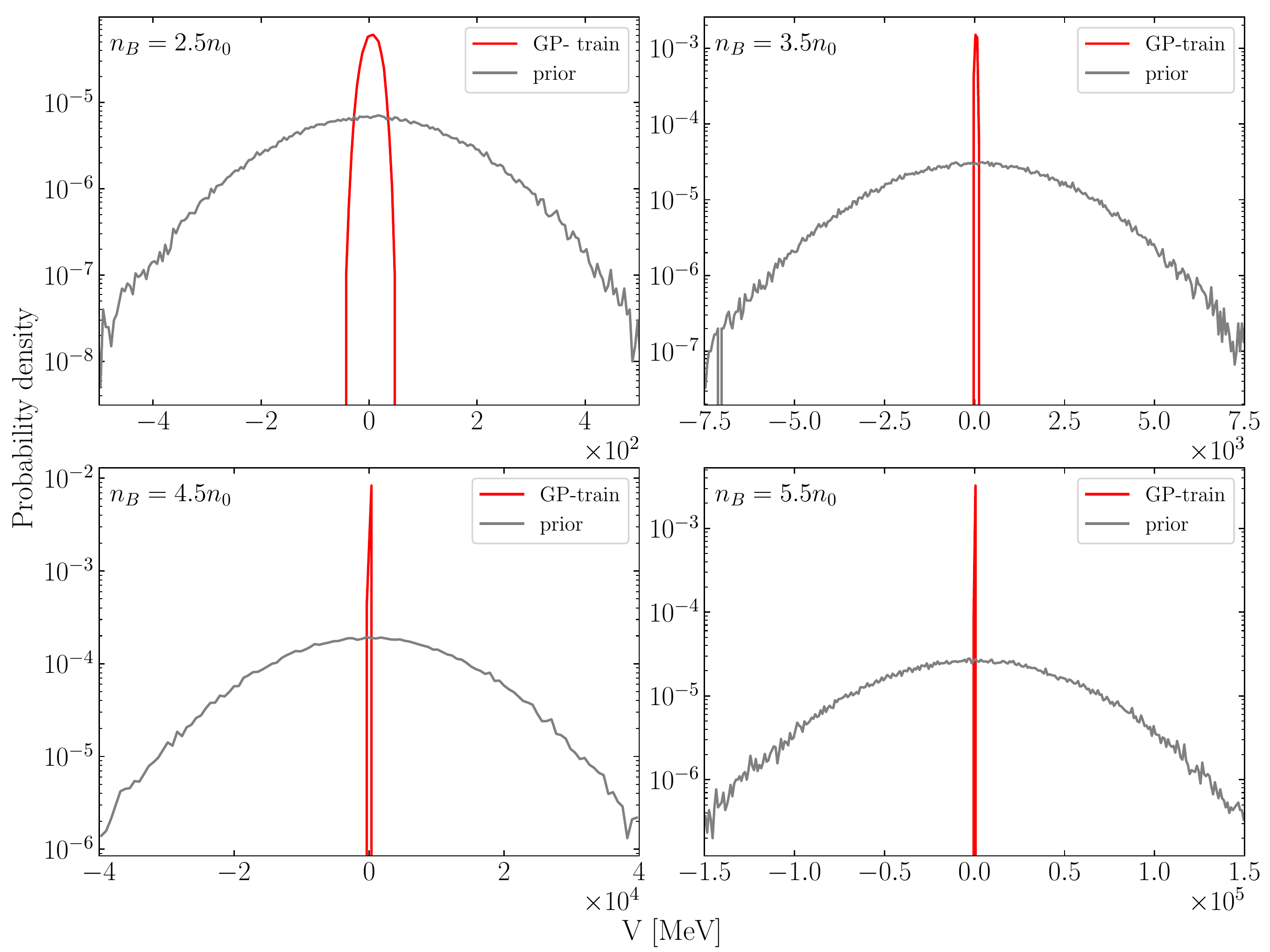}
   \caption{(Color online) Distribution of the potential $V(n_B)$ at four different values of the baryon density. Compared are the prior used in the Bayesian inference (grey lines) and the distributions of the constrained potentials used in training the GP models (red lines).}
   \label{GP_vs_prior}
 \end{figure*}

Regarding the prior for the Bayesian inference, it is important to note that a prior based only on the GP training constraints could also be a good starting point for the parameter estimation but not a necessary one. The physics constraints can disfavor the acausal range for the parameters. However, we employ this range only as a soft constraint in the prior as we use the mean and width of each coefficient independently, thereby the prior is not limited by the correlations between the coefficients from the GP-training set. This results in inferred potentials which can also be outside the training range for the Gaussian Process models. In fact, the range of physically constrained potentials used in training the GP models is only a small fraction of the  prior distribution of the potentials at any given density as shown in figure \ref{GP_vs_prior}.  Here, the prior distributions of the potential V at four different densities (grey lines) is compared to the range of the potential used in training the GP models.

It is true that the predictions of GP models for potentials very far from its training range may not be reliable. However, in this case, the prediction uncertainty given by the GP models ($\sigma_{i,GP}$) will also be very high. This would then result in a very low likelihood

\begin{equation}\ln P(\textbf{D}|\boldsymbol\theta)=-\frac{1}{2}\sum_i \left[\frac{( x^{\boldsymbol\theta}_i - d_i)^2}{\sigma_i^{2}}+(\ln(2\pi\sigma_i^{2}))\right]\label{likelihood}\end{equation}

as the uncertainty term takes into account the uncertainty in the experiment as well as in the GP predictions  (${\sigma_{i}^2=\sigma_{i,exp}^2+\sigma_{i,GP}^2}$).  Hence, one can consider the GP training data as the lower bound of the “effective prior” seen by the MCMC for the Bayesian inference and it is possible for the MCMC to sample potentials outside this range if evidence demands.

\section{The closure tests}\label{secclose}
\begin{figure*}[t]
   \includegraphics[width=0.85\textwidth]{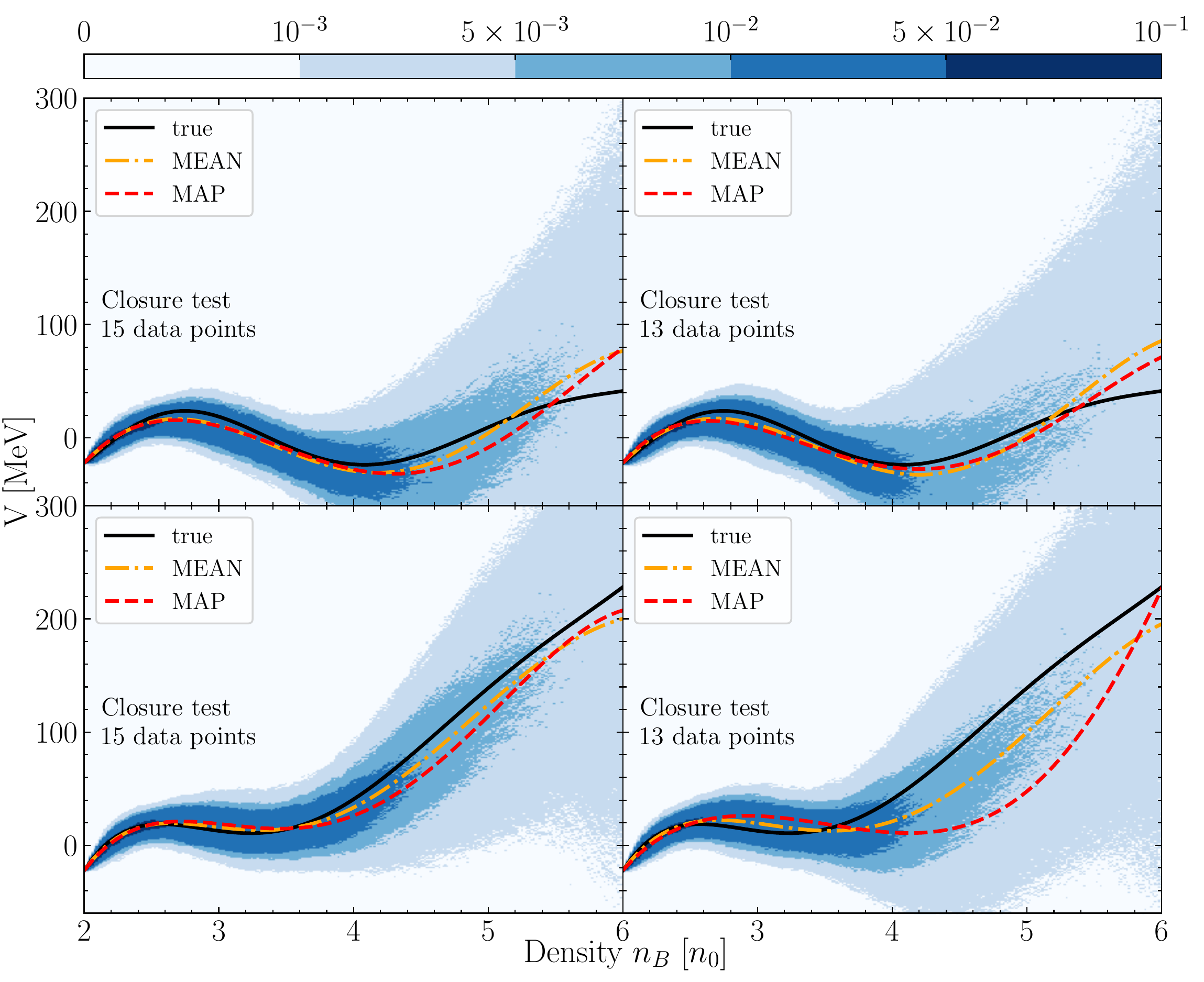}
   \caption{(Color online) Visualisation of the posterior constructed in the closure tests. The 'ground-truth' EoS is plotted as black solid line.  The red dashed and orange dot-dashed curves are the MAP and MEAN EoS respectively. Each row in the figure corresponds to the posterior for a random "ground-truth" EoS. The plots in the first column shows the posterior constructed using all 15 observables and the posterior constructed using 13 observables is shown in the second column. The $\left\langle m_T \right \rangle - m_0$ values for $\sqrt{s_{\mathrm{NN}}}$= 3.83 and 4.29 GeV were removed in the test results shown in the second column.}
   \label{close}
 \end{figure*}

Two different closure tests are performed to verify the capability of the Bayesian inference method in constraining the EoS and the sensitivity of the method to the choice of experimental data. In these tests, a random EoS is assumed to be the 'ground-truth' EoS and the UrQMD predictions of $v_2$ and $\left\langle m_T \right \rangle - m_0$ for this 'ground-truth' EoS are then taken as experimental observations with the experimental uncertainties. The posterior distribution of EoSs is then constructed via the MCMC sampling using these observations. By comparing the reconstructed posterior with the 'ground-truth' EoS, we can infer the ability and accuracy of the method to reconstruct the EoS. To study the sensitivity of the results on our choice of experimental data, the same test is repeated without using the $\left\langle m_T \right \rangle - m_0$ for $\sqrt{s_{\mathrm{NN}}}$= 3.83 and 4.29 GeV. Additionally, the MEAN and MAP EoSs are also compared against the 'ground-truth' EoS to validate the reliability in extracting a 'most probable EoS'.  The results of these tests are visualised in figure \ref{close}.

The tests reveal that the bayesian inference technique we use can well constrain the high density EoS using $v_2$ and $\left\langle m_T \right \rangle - m_0$ values for beam energies $\sqrt{s_{\mathrm{NN}}}$= 2- 10 GeV, assuming that all experimental observables are simulated consistently. While using all 15 observables, the extracted MEAN and MAP EoSs closely match the "ground-truth" EoS for densities up to 6 $n_0$. In this case, the EoS is well constrained for densities up to 4 $n_0$ and for densities 4- 6 $n_0$, the posterior distribution has large variance. However, when the two data points are removed from the observables, the MEAN and MAP EoSs extracted in this case may not always represent the ground-truth accurately. In the first example (figure \ref{close}, top right plot), the  MEAN and MAP EoSs closely match the ground-truth for densities up to 5 $n_0$. However, in the second example (figure \ref{close}, bottom right plot), the MAP and MEAN EoS deviates from ground truth for densities above 3.5 $n_0$. Nevertheless, the overall trend of the MEAN and MAP EoSs and the posterior distribution doesn't vary drastically even if $\left\langle m_T \right \rangle - m_0$ values for $\sqrt{s_{\mathrm{NN}}}$= 3.83 and 4.29 GeV are not used in the inference procedure. This is indicative of the fact that if the observations are consistent with each other, removing few observations from the evidence wouldn't affect the extracted posterior distribution though this could lead to larger variance in the posterior distribution.

 \section{Prior vs posterior}\label{secpripos}
\begin{figure}[]
   \includegraphics[width=0.5\textwidth]{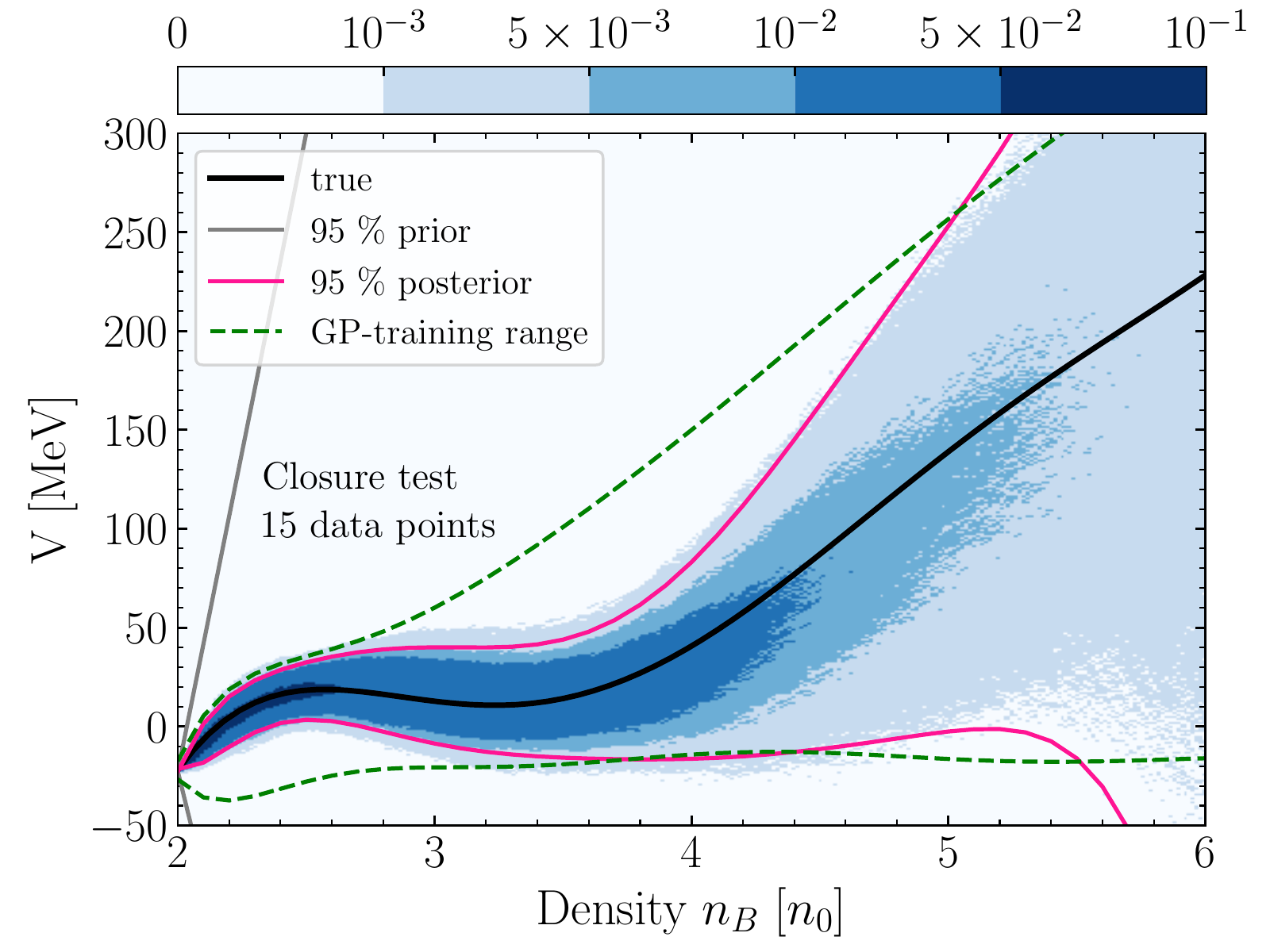}
   \caption{(Color online) Visualization of prior (dark grey) and posterior (magenta) ranges together with the GP-training range (green dashed) for one of the closure tests with simulated data. The prior range is much broader than the GP-training range which is again broader than the posterior.}
   \label{close_prior_vs_post}
 \end{figure} 
   \begin{figure}[]
   \includegraphics[width=0.5\textwidth]{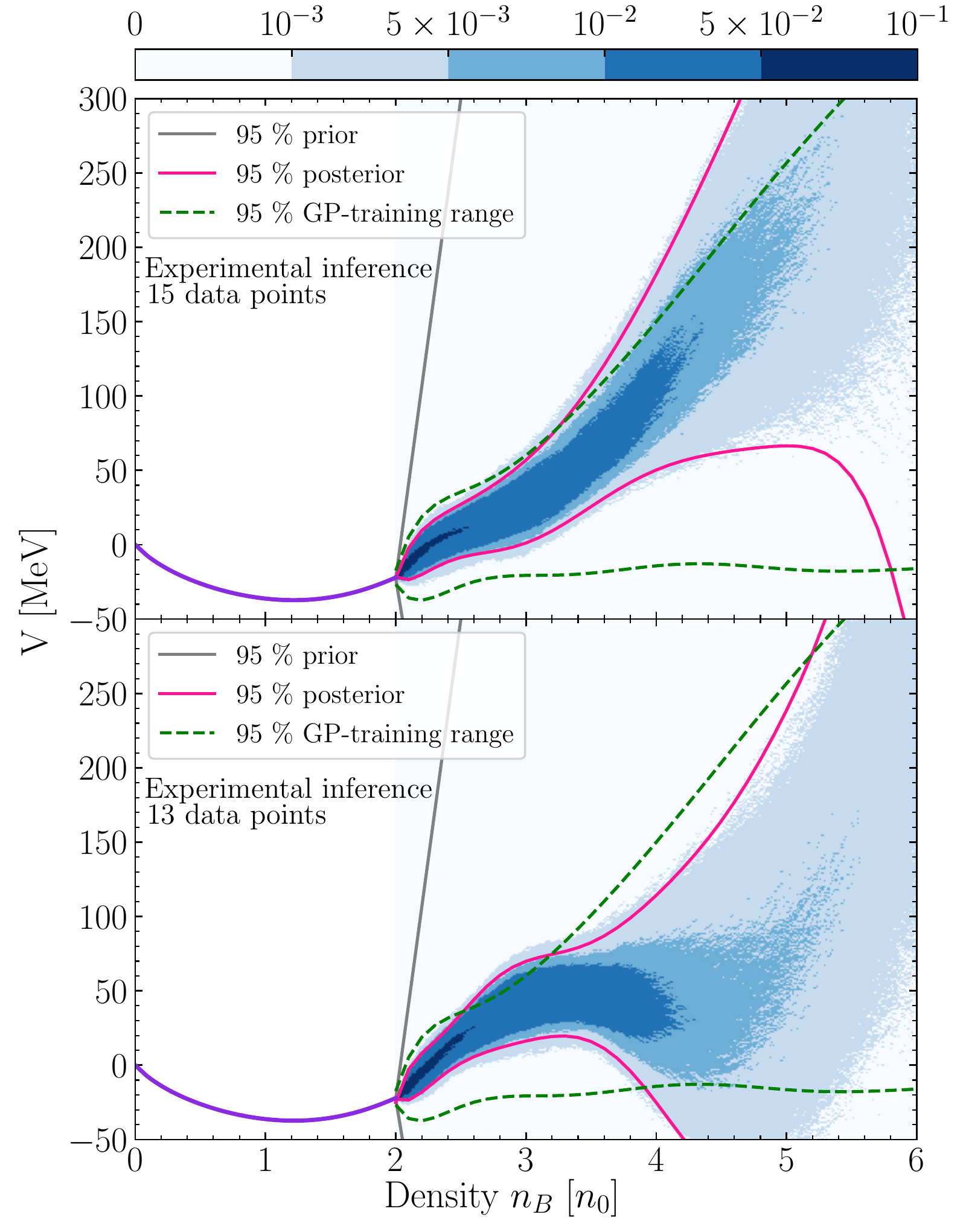}
   \caption{(Color online) Same as figure \ref{close_prior_vs_post} but with the real data.}
   \label{data_prior_vs_post}
 \end{figure} 

A comparison of the prior and the posterior distribution is essential to understand how much information is gained from the data. However, as already shown in figure \ref{GP_vs_prior}, the actual prior distribution used is extremely broad. Nevertheless, we have visualized the prior and posterior distributions (95$\%$ confidence intervals, grey and magenta lines), together with the GP training range (green dashed lines), for the potential in one of the closure tests, together in one single plot in figure \ref{close_prior_vs_post}. As one can see, the actual prior is much broader than the posterior for the closure test with the simulated data. This is also true for the real data as shown in figure \ref{data_prior_vs_post}. Thus, it is clear that there is a significant information gain.

 \section{Sensitivity to high beam energies}\label{secsense}
 \begin{figure}[]
   \includegraphics[width=0.5\textwidth]{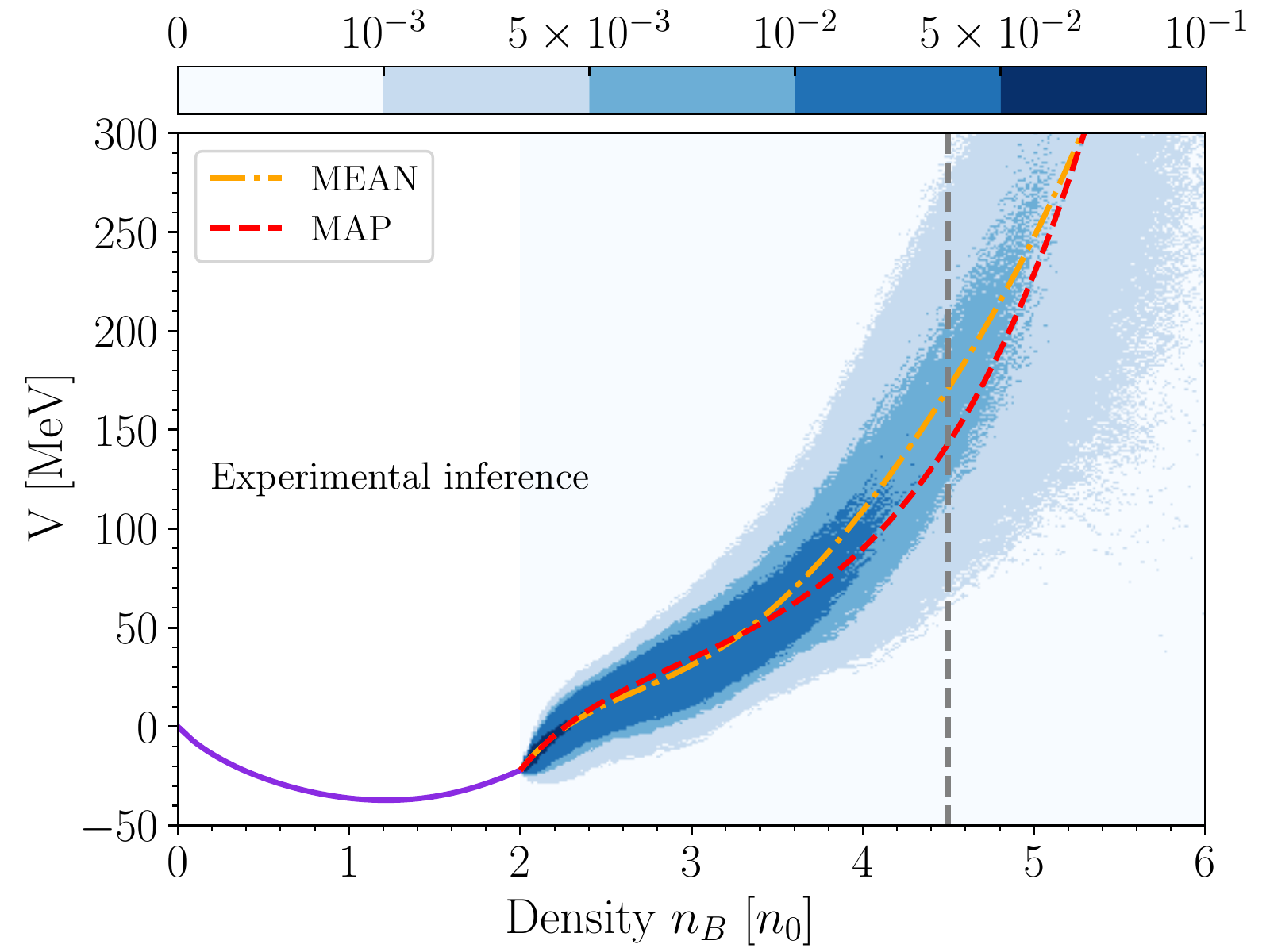}
   \caption{(Color online) The probability density functions of the extracted posterior as a function of density, for a scenario where the $\langle m_T\rangle -m_{0}$ of the two highest beam energies is removed from the analysis.}
   \label{rem_highmt}
 \end{figure}
To check whether it is really the two low energy $\langle m_T \rangle -m_{0}$ data points which are most relevant in the inference, the bayesian inference was performed after removing the $\langle m_T \rangle -m_{0}$ data points at two highest beam energies .
It was found that the resulting constraints are less sensitive to removing data points from higher beam energies (or higher densities). This can be seen in figure \ref{rem_highmt}  which shows the probability density functions of the extracted posterior as a function of density, for a scenario where the mean $\langle m_T \rangle -m_{0}$ of the two highest beam energies (7.7 GeV and 8.865 GeV) is removed from the analysis. The resulting potential is very similar to that with the two points which supports our statement of less sensitivity to the data at the highest beam energies. It is therefore clear that the constraints on the EoS are very sensitive on the two data points at low beam energies and much less sensitive on the high energy points, within the beam energy range currently under consideration.

\section{Differential spectra}
\label{secdiff}
As mentioned in the letter, the extracted EoS can be tested with various observables from heavy ion collisions. Several recent works have explored other different observables that are sensitive to the equation of state \cite{Savchuk:2022aev,Li:2022iil,Savchuk:2022msa,Reichert:2023eev} (for example pion HBT, dilepton production, net proton fluctuations). In the current work we put an emphasis on the integrated values of the mean transverse momentum and elliptic flow, as these can be calculated with a limited amount of simulated events. The main restricting factor for our analysis is the computational effort required to simulate the different observables for the various equations of state required to train the Gaussian Process emulator. Once the EoS is constrained, of course, many observables for many beam energies and system sizes can be predicted and compared. We are also planning to make the model available in the future so that all these possibilities can be explored.

\begin{figure}[]
   \includegraphics[width=0.49\textwidth]{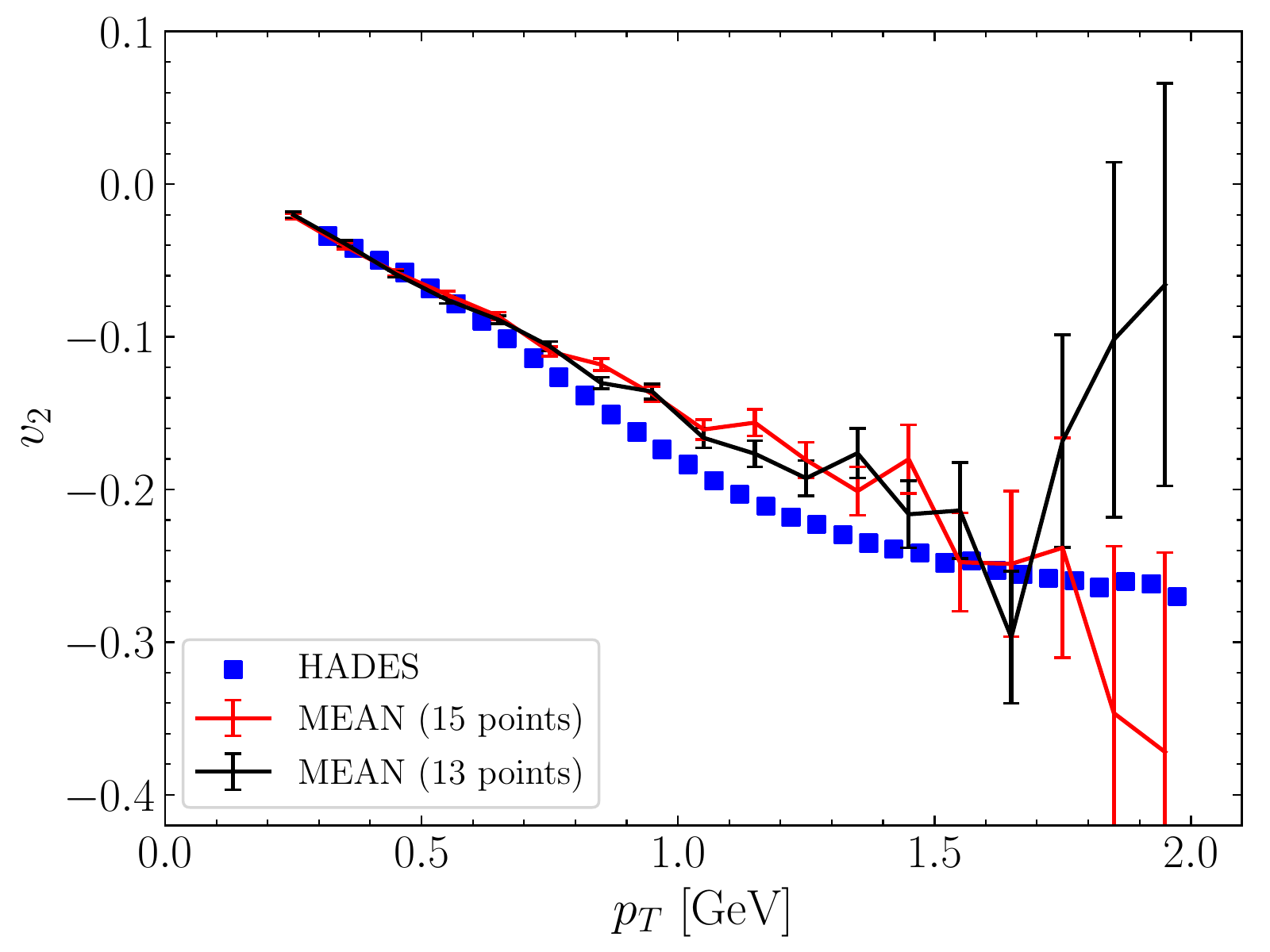}
   \caption{(Color online) Differential elliptic flow of protons for mid-central collisions of AuAu at $E_{\mathrm{lab}} = 1.23 A$ GeV. HADES data are compared to simulations with the two different MEAN equations of state.}
   \label{diffv2}
 \end{figure}

In addition to the directed flow, which was shown in the letter, a comparison with recently published HADES data on the differential elliptic flow in Au-Au collisions at $E_{\mathrm{lab}} = 1.23 A$ GeV \cite{HADES:2022osk} is presented here. This comparison of the two different MEAN EoS to HADES data is shown in figure \ref{diffv2}. As one can see, the extracted EoSs reproduce the $p_T$ dependence nicely up to a proton momentum of 1 GeV. Above this range, the model slightly overestimates the elliptic flow compared to HADES data. The reason for this is likely a small momentum dependence of the potential interaction which is not considered in the present approach. It is however important to note that the integrated elliptic flow is only sensitive to the flow around the maximum of the proton $p_T$ distribution which corresponds roughly to $p_T$ between 300 and 400 MeV.

\end{document}